\documentclass[twocolumn,superscriptaddress,aps,a4,prl]{revtex4-2}
\usepackage{graphicx} 

\usepackage[english]{babel}
\usepackage{color}
\usepackage{feynmf}
\usepackage{amsmath,amssymb}
\usepackage{verbatim}
\usepackage{xcolor}
\usepackage{physics}
\usepackage{dcolumn}
\usepackage[a4paper,margin=.7in]{geometry}
\usepackage[normalem]{ulem}

\usepackage{makecell}
\usepackage{amsmath,amssymb}
\usepackage{comment}

\usepackage[colorlinks=true, allcolors=blue]{hyperref}
\usepackage{booktabs}
\usepackage{cellspace}

\begin{document}

\title{Stringent Constraints on New Pseudoscalar \& Vector Bosons
from \\ Precision Hyperfine Splitting Measurements}
\author{Cedric Quint}
\affiliation{Max Planck Institute for Nuclear Physics,\\
 Saupfercheckweg 1, 69117 Heidelberg, Germany}
  \author{Fabian Hei\ss e}
 \affiliation{Max Planck Institute for Nuclear Physics,\\
 Saupfercheckweg 1, 69117 Heidelberg, Germany}
 \author{Joerg Jaeckel}
 \affiliation{Institute for Theoretical Physics,\\
 Philosophenweg 16, 69120 Heidelberg, Germany}

 \author{Lutz Leimenstoll}
  \affiliation{Institute for Theoretical Physics,\\
 Philosophenweg 16, 69120 Heidelberg, Germany}
\author{Christoph H. Keitel}
\author{Zolt\'an Harman}
\affiliation{Max Planck Institute for Nuclear Physics,\\
 Saupfercheckweg 1, 69117 Heidelberg, Germany}
 
\begin{abstract}

Axion-like particles and similar new pseudoscalar as well as vector bosons coupled to nucleons and electrons are predicted to lead to spin-dependent forces in atoms and ions. We argue that hyperfine structure measurements in  hydrogen- and lithium-like charge states are a sensitive probe to this effect.
Employing specific differences of these splittings
reduces uncertainties due to nuclear effects in hyperfine structure calculations and measurements.   
Using this, we show that existing measurements on Be provide competitive limits in the region $m_{\phi}\gtrsim 100\,{\rm keV}$, confirming, or improving by up to a factor of 2, existing constraints for pseudoscalar couplings, depending on the nuclear model. We also find that future measurements on Cs have a further factor of $2-2.5$ improved discovery potential for pseudoscalars and an order of magnitude for new vector bosons when compared with the corresponding current constraints. 
\end{abstract}

\maketitle

In recent years, there has been increasing interest in investigating physics beyond 
the standard model (SM) using the tools of precision atomic spectroscopy (cf.\ Refs.\ \cite{Karshenboim_2002, Jaeckel_2010, PhysRevD.82.113013, PhysRevA.95.032505, DEBIERRE2020135527, ADKINS20221,   Sailer2022, potvliege2025spectroscopylightatomsbounds, Door_2025, Cong:2025wge, PhysRevA.108.052825}) as well as similar approaches (cf.~Refs.~\cite{Wei:2022mra, PhysRevLett.130.133202, Beyer:2023bwd, wilzewski2024nonlinearcalciumkingplot, xu2025newconstraintsaxionmediated,  elder2025prospectsdetectingnewdark, PhysRevLett.134.055001}). Notably, a recent review~\cite{Cong:2024qly} highlighted the sensitivity of atomic systems to new physics-induced spin-dependent interactions.  
This motivates
new physics searches enabled by the prediction that the exchange of new bosons also implies the existence of new forces and potentials~\cite{Moody:1984ba,Dobrescu:2006au}. 

Most existing constraints focus on light atoms or isotope shifts, where theoretical and experimental precision exceeds that of heavier-element spectroscopy \cite{wilzewski2024nonlinearcalciumkingplot, DEBIERRE2020135527}. We argue that highly charged ions can also play an important role, notably for pseudoscalars and vector bosons.

Generally speaking, the experimental approach best suited for investigating forces induced by new particles depends on the Lorentz structure of the interaction~\cite{Moody:1984ba,Dobrescu:2006au}. In this Letter, we focus on pseudoscalar and vector 
interactions, leading to a 
dipole-dipole interaction between the electron(s) and the nucleus. 

Pseudoscalars appear in a wide range of extensions to the SM. Light variants with interactions that  are suppressed by a large new physics scale are often  called axions or axion-like particles (ALPs)~\cite{Kim:1986ax,Jaeckel:2010ni,Beacham:2019nyx,Agrawal:2021dbo}. This is motivated by their similarity to QCD axions that arise as a consequence of the Peccei-Quinn (PQ) solution to the strong charge parity (CP) problem in QCD \cite{PhysRevLett.38.1440,Weinberg:1977ma,Kim:1979if,Shifman:1979if,Zhitnitsky:1980tq,Dine:1981rt} (see~\cite{DiLuzio:2020wdo} for an overview of models). In field theory, such pseudoscalars arise as pseudo Nambu-Goldstone bosons of a spontaneously broken global symmetry. 
They can also arise in string theory scenarios \cite{Conlon:2006tq, Conlon:2006ur, Arvanitaki:2009fg, Cicoli:2012sz, Marsh:2015xka, Acharya:2015zfk,Visinelli:2018utg, Broeckel:2021dpz, Demirtas:2021gsq,Mehta:2021pwf,Gendler:2023kjt,Gendler:2024adn,Petrossian-Byrne:2025mto,Loladze:2025uvf}.
Similarly, extra vector bosons can be viewed as one of the simplest extensions of the SM, notably its gauge symmetries by a U(1) factor~\cite{Jaeckel:2012mjv,Fabbrichesi:2020wbt}. They can interact directly (see, e.g.~\cite{He:1990pn,He:1991qd,Heeck:2011wj,Bauer:2018onh} for some flavor dependent options), via kinetic mixing~\cite{Okun:1982xi, Holdom:1985ag,Foot:1991kb} or higher dimensional operators~\cite{Dobrescu:2004wz,Dobrescu:2006au}, and they, too, are a generic feature of string based models~\cite{Goodsell:2009pi,Goodsell:2009xc,
Bullimore:2010aj, Cicoli:2011yh, Camara:2011jg, Marchesano:2014bia, Anastasopoulos:2020xgu, Hebecker:2023qwl,
Sheridan:2024vtt,Coudarchet:2025dfd}.

Both pseudoscalars and light vectors  
arise in solutions to various open problems in physics, most notably as a possible light dark matter (DM) candidate~\cite{Preskill:1982cy,Abbott:1982af,Dine:1982ah,Nelson:2011sf,Arias:2012az,Marsh:2015xka, adams2023axiondarkmatter,Antypas:2022asj}.

Due to their Lorentz structure, interactions mediated by such particles induce spin-spin couplings in atomic systems~\cite{Moody:1984ba,Dobrescu:2006au}. Hence, they give rise to an additional contribution to the hyperfine splitting (HFS) in such systems, thereby giving access to possible experimental signatures. 

For higher masses, such interactions are expected to be short-ranged, making them harder to detect in fifth force experiments, and limits are typically weaker~\cite{Cong:2024qly}. To access the $\sim (1-100)\,{\rm MeV}$ range, 
choosing an element with a high nuclear charge $Z$ is an intuitive step, resulting in a shorter mean distance between the bound electron and the nucleus. However, finite nuclear size (FNS) effects grow with increasing $Z$, reducing theoretical precision. We mitigate this by considering specific differences of hyperfine splittings. These are weighted differences between the hyperfine splitting energy of lithium- and hydrogen-like ions, with the weighting factor defined such that the effects due to the magnetization distribution of the nucleus are largely suppressed  \cite{PhysRevLett.86.3959}. The study of HFS in both charge states has recently been enabled by experimental developments with stored and trapped ions in the high-$Z$ \cite{10.1038/ncomms15484} and low-$Z$ \cite{Dickopf_2024} regimes. This improves HFS bounds on pseudoscalar and vector couplings, even for hydrogen and helium \cite{Fadeev_2019,Cong:2024qly}. However, this also results in a suppression of the new physics effects for hydrogen and helium. This suppression is less pronounced for medium to highly charged ions (HCI) that can thus yield competitive bounds on those new interactions: as we show below, in the high-mass regime, the bounds can be improved by up to a factor of 2.5 for pseudoscalars, and an order of magnitude for vectors compared to the best previous results. Another advantage of HCIs is that, via the choice of the isotope, one can selectively constrain proton and neutron coupling constants \cite{zyb6-lvy8} opening promising directions for further improvement.

{\it Theory.\ }-- The hyperfine structure of hydrogen has a rich history in the study of relativistic and radiative corrections to the energy levels predicted by the Schr\"odinger equation. 
This field has reached a high level of maturity, enabling remarkable precision in theory and experiment~\cite{PhysRevA.52.3686, PhysRevA.56.252, 10.1038/ncomms15484, Dickopf_2024, Schneider2022, PhysRevResearch.2.013364}. 
The development of diverse methods in recent decades has yielded numerous insights into HFS in hydrogen- and lithium-like ions. The former can be encapsulated into the following parametrization of the ground-state HFS of hydrogen-like ions~\cite{PhysRevA.52.3686}
\begin{eqnarray}
    \triangle E_{\text{H-like}} &=& \frac{4}{3}\alpha(\alpha Z)^3 \frac{\mu}{\mu_N} \frac{m_e}{m_p}\frac{2I+1}{2I} \frac{m_e}{(1+\frac{m_e}{M})^3}  \nonumber \\
    &\times& (A_{1s}(\alpha Z)(1-\delta_{1s})(1-\varepsilon_{1s}) + x^{1s}_{\text{rad}}).
\end{eqnarray}
Here, $\alpha$ is the fine-structure constant, $Z$ the nuclear charge, $m_e$ ($m_p$) the electron (proton) mass, $\mu$ ($\mu_N$) the nuclear magnetic moment (magneton), $I$ the nuclear spin, and $M$ the nuclear mass. We use natural units ($\hbar=c=1$).
The factor $A_{1s}(\alpha Z)$ denotes the relativistic correction~\cite{PhysRev.35.1447}.
Furthermore, $\delta_{1s}$ is the nuclear charge distribution correction, and $\varepsilon_{1s}$ the nuclear magnetization distribution correction (Bohr-Weisskopf (BW) correction). Finally, $x^{1s}_{\text{rad}}$ denotes radiative corrections ~\cite{PhysRevA.52.3686, PhysRev.121.1128, PhysRevLett.51.985}).\par

For lithium-like ions, one can observe 
a similar structure in HFS contributions, with the addition of interelectronic interaction contributions~\cite{PhysRevA.52.3686}
\begin{eqnarray}
    \triangle E_{\text{Li-like}} &=& \frac{1}{6}\alpha(\alpha Z)^3 \frac{\mu}{\mu_N} \frac{m_e}{m_p}\frac{2I+1}{2I} \frac{m_e}{(1+\frac{m_e}{M})^3}  \nonumber \\
    &\times& \big(\big[A_{1s}(\alpha Z)+\frac{1}{Z}B(\alpha Z) + \dots\big]\nonumber \\
    &\times& (1-\delta_{2s})(1-\varepsilon_{2s}) 
    + x^{2s}_{\text{rad}}\big).
\end{eqnarray}
Here, $A_{2s}(\alpha Z)$ denotes the relativistic factor for the $2s$ system, $B(\alpha Z)$ denotes the photon exchange contribution to first order, and $x^{2s}_{\text{rad}}$ parametrizes the radiative corrections for the lithium-like system. Moreover, $\delta_{2s}$ and $\varepsilon_{2s}$ again denote the nuclear charge distribution and BW correction, respectively. We note that the photon exchange contributions admit a $1/Z$ expansion, since each exchange provides a factor $\alpha$ in the $Z\alpha$ expansion.
We calculate the energy shift induced by the spin-dependent interaction between an electron and the nucleus. To model nuclear couplings of new particles, we use the single-particle Schmidt model~\cite{SchmidtModel}. Here, the spin of a nucleus with an odd number of nucleons arises from the total angular momentum of the valence nucleon. All other nucleons do not contribute to the spin-dependent coupling. Hence, the electron effectively couples to that valence nucleon only \cite{leimenstoll, SchmidtModel}.
We therefore interpret the interaction as a coupling to the dipole operator $\pmb{\mu}_N = g^{\text{NP}}_I \pmb{I}$, where $g^{\text{NP}}_I$ is the new physics associated nuclear $g$-factor evaluated using nuclear shell states. 
Although reasonably well-justified, this assumption could be improved upon \cite{Kimball_2015}.
We account for the introduced error by adopting conservative constraints based on the nuclear model comparison in ~\cite{Kimball_2015} (see the Appendix for details).\par 
We represent the corresponding interaction Lagrangian as
\begin{align}
    \mathcal{L}_{\text{int}} = \sum_{k = e,N} ig_{k}\bar{\psi}_k\gamma^5 \psi_k \phi + y_{k}\bar{\psi}_k\gamma^{\mu} \psi_k Z_{\mu}  .
\end{align}
Here, $\psi_k$ denotes the Dirac field of the electron $e$ or nucleon $N \in \{n,p\}$, where $n$ denotes the neutron, and $p$ the proton. Moreover, $g_{k}/y_k$ is the respective coupling constant of the pseudoscalar/vector field to the indexed particle,  $\phi/Z_{\mu}$ the pseudoscalar/vector field, and $\gamma^5/\gamma^{\mu}$ denote the Dirac gamma matrices. 
Using 
the non-relativistic approximation 
for the dynamics of the nucleons, one obtains the following potential from the aforementioned interaction Lagrangian \cite{Fadeev_2019,Dzuba_2018}

\begin{eqnarray}
    V_{\phi/Z}(r) &=& \frac{g_Ng_e}{8\pi m_N} \pmb{\mu}_N \cdot \pmb{T} \frac{e^{-m_{\phi/Z}r}}{r}(m_{\phi/Z}+1/r) \nonumber \\
    \pmb{T} &=& \begin{cases}
        i\hat{\pmb{r}}\gamma^0\gamma^5, \,\, &\text{Pseudoscalar} \\
        \hat{\pmb{r}} \times \gamma^0\pmb{\gamma}, \,\, &\text{Vector}
    \end{cases}.\label{eq:vphi}
\end{eqnarray}

Here, $r$ is the radial variable in the system of the nucleus, $\pmb{\mu}_N$ the aforementioned dipole operator as described in the Appendix, $m_N$ and $m_{\phi/Z}$ are the masses of the nucleon and the exchange boson, respectively,  and $\hat{\pmb{r}}=\pmb{r}/r$ the unit coordinate vector.

Using the Wigner–Eckart theorem \cite{10.1119/1.1937653}, we obtain the induced HFS

\begin{align}
    \triangle E &=  g_I^{\text{NP}}\frac{g_Ng_e}{4\pi} \frac{m_e^2}{2m_N} \frac{(2I+1)}{\sqrt{6}} \nonumber \\ 
    &\times \int_{\pmb{r}} \psi^{\dagger}_{n,\kappa,m}(\pmb{r})(\beta(r)\pmb{T})\psi_{n,\kappa,m}(\pmb{r})\, \text{d}\pmb{r}. \label{eq:shift}
\end{align}
Here, $\kappa$ is the relativistic angular momentum quantum number of the electron. The functions $\psi_{n,\kappa,m}$ called Dirac-Coulomb wavefunctions are obtained in the nuclear potential corresponding to the Fermi proton distribution (see~\cite{2006sham.book.....D,10.1119/1.1937653} and the Appendix).
Moreover, we defined the potential function
\begin{equation}
    \beta(r) = \frac{e^{-m_{\phi}r}}{r}(m_{\phi}+1/r).
\end{equation}
From this, the new physics induced hyperfine structure becomes readily calculable in one electron systems. 
    For lithium-like ions, however, interelectronic interactions become relevant due to the overlap of the core and valence $s$ electrons in the nuclear region. We treat this effect as discussed in ~\cite{PhysRevResearch.2.013364}, i.e., via the addition of screening terms to the nuclear potential and perturbation theory in $\alpha$. The former is known as treating the system within the \textit{extended Furry picture}~\cite{PhysRevResearch.2.013364, PhysRevA.63.032506, ORESHKINA2008675}. To this end, we take the effective potential for the electrons in the lithium-like system to be the Kohn-Sham (KS) potential~\cite{PhysRev.140.A1133, PhysRev.136.B864} from density functional theory~\cite{ORESHKINA2008675, PhysRevA.67.022512}, as explicitly given in the Appendix. The KS potential is determined by the electron probability density around the nucleus, which we obtain self-consistently from Dirac-Hartree-Fock (DHF) wavefunctions~\cite{Johnson}. The hyperfine structure calculations are then carried out using radial wavefunctions that were obtained by solving the Dirac equation with this modified potential term. \par
Up to first order in $\alpha$, we consider the one photon exchange contribution to the HFS in the lithium-like system perturbatively. Schematically, this contribution is given by the following diagram:
\begin{center}
\begin{equation}
\begin{gathered}
\begin{fmffile}{photonexchange}
\begin{fmfgraph*}(70,40)
    \fmfstraight
    \fmfleft{i1,i2,i3} 
    \fmfright{o1,o2,o3}
    \fmffreeze
    \fmf{phantom}{i1,v11,v12,o1}
    \fmf{dbl_plain}{i2,v21,v22,o2} 
    \fmf{dbl_plain}{i3,v31,v32,o3}
   
    \fmffreeze

    \fmf{photon}{v21,v31}
    \fmf{dashes}{v22,v12}
    \fmfv{decor.shape=square,decor.size=.15cm}{v12}
\end{fmfgraph*}
\end{fmffile}
\end{gathered}
\quad
+
\quad
\begin{gathered}
\begin{fmffile}{counterterm}
\begin{fmfgraph*}(70,40)
    \fmfstraight
    \fmfleft{i1,i2,i3} 
    \fmfright{o1,o2,o3}
    \fmffreeze
    \fmf{phantom}{i1,v11,v12,o1}
    \fmf{dbl_plain}{i2,v21,v22,o2} 
    \fmf{phantom}{i3,v31,v32,o3}
   
    \fmffreeze

    \fmfv{decor.shape=circle,decor.size=.2cm,decor.fill=empty}{v21}
    
    \fmf{dashes}{v22,v12}
    \fmfv{decor.shape=square,decor.size=.15cm}{v12}
\end{fmfgraph*}
\end{fmffile}
\end{gathered},
\end{equation}

\end{center}
where a double line denotes an electron propagating in the screened nuclear potential $V$, the wavy line represents the exchange of a photon, and the dashed line terminated by a square the pseudoscalar/vector potential $V_{\phi/Z}$ in Eq.~(\ref{eq:vphi}). 
The second diagram denotes a counter-term contribution we add to avoid double-counting interaction effects when calculating the photon exchange contribution within the extended Furry picture. Hence, the white circle denotes an insertion of the screening potential. For details on this calculation, 
see the Appendix and Ref.~\cite{PhysRevResearch.2.013364}.\par
With this approach, we have included Coulomb interactions between electrons to high order by using a self-consistent screening potential and 
included relativistic corrections up to first order by including photon exchange. \par

We derive constraints from the \emph{specific difference} of HFS energies between lithium- and hydrogen-like ions of the same isotope. This weighted difference is constructed to suppress uncertainties originating from the BW effect~\cite{PhysRev.77.94}. \par 
These specific differences were first discussed for bismuth in Ref.~\cite{PhysRevLett.86.3959}, and can be calculated with significantly higher precision than the individual HFS contributions. They are defined as
\begin{equation}
    \triangle' E = \triangle E_{\text{Li-like}} - \xi \triangle E_{\text{H-like}}.
\end{equation}
Here, the factor $\xi$ is the ratio of the BW shifts, i.e., 
\begin{equation}
    \xi = \frac{\triangle E^{\text{BW}}_{\text{Li-like}}}{\triangle E^{\text{BW}}_{\text{H-like}}}.
\end{equation}
Since $\xi$ can be determined more precisely than individual BW terms, the specific difference yields a robust observable with reduced uncertainty. Note that $\xi$ is fully determined within the SM and FNS effects enter only as small corrections to the new physics contribution. We have ensured that this is a small effect.

For many elements, tabulated values of $\xi$ are unavailable. In these cases, we extract $\xi$ from the BW shifts in ~\cite{Indelicato}.
Where available, we use dedicated calculations of $\xi$. Our bounds are insensitive to variations in $\xi$ beyond the fourth digit.\par 
We compare with constraints from hydrogen and helium. For these elements, an alternative specific difference $D_{21}$ can be defined \cite{Karshenboim_2002}. It compares the HFS of a $1s$ and excited $2s$ state 
\begin{equation}
    D_{21} = 8 \triangle E_{2s} - \triangle E_{1s},
\end{equation}
where $\triangle E_{1s}$ is the $1s$ HFS, and $\triangle E_{2s}$ corresponds to the respective $2s$ HFS. This is analogous to $\triangle’ E$ with $\xi=1/8$, which approximately holds for low-$Z$ elements.
In Table~\ref{tab:table}, we present the theory and experimental values for $D_{21}$ in helium and hydrogen along with the values for $\triangle' E$ in heavier elements. Details on the experimental and theoretical precision are given in the Appendix.\par

\begin{figure*}

\centering
\includegraphics[width=.95\textwidth]{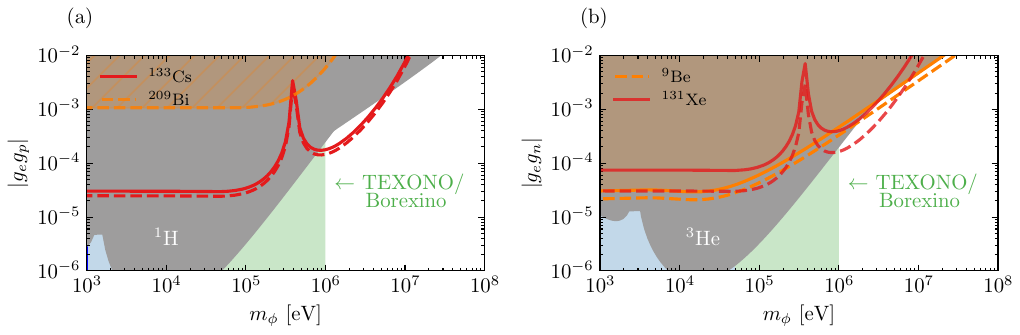}\\
\includegraphics[width=.95\textwidth]{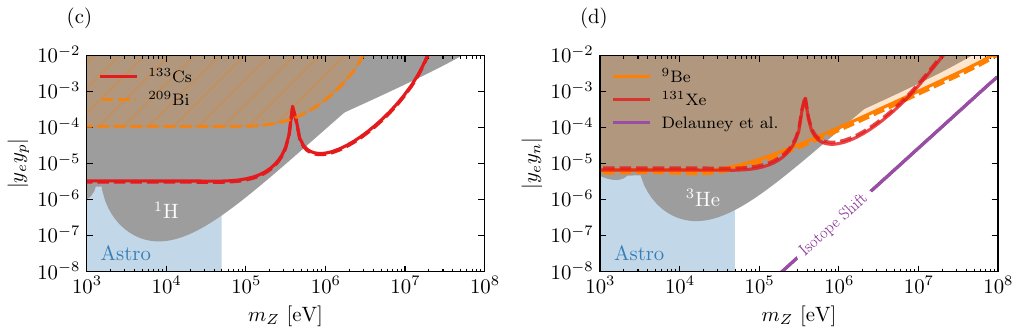}
\caption{Bounds on the pseudoscalar (top) and vector (bottom) parameter space. The left plots $(a)$ and $(c)$ show bounds on the product of the electron \& proton couplings, whereas the right, $(b)$ and $(d)$, show the bounds on the product of electron \& neutron couplings. Red lines (projections) and orange regions (current constraints) are derived in this work. The gray area on the left denotes bounds from the hydrogen specific difference $D_{21}^{\rm H}$ \cite{Cong:2024nat} and $1s$ HFS \cite{PhysRevA.108.052804}. The gray area on the right-hand side denotes bounds from the helium-$3$ specific difference $D_{21}^{\rm He}$ and $1s$ HFS. In the vector electron–neutron case, isotope-shift constraints are shown in purple \cite{Cong:2024qly, PhysRevD.96.115002}.
Green shows TEXONO~\cite{Chang_2007} and Borexino~\cite{BOREXINO:2025dbp} constraints; blue denotes astrophysical constraints \cite{Cong:2024qly}. Solid lines show conservative bounds accounting for estimated uncertainties (see Appendix) and possible Schmidt-model errors~\cite{Kimball_2015}. Dashed lines indicate Schmidt model results. The bismuth constraint is a projection assuming that a direct measurement of the magnetic moment will agree with the NMR result.}  
\label{fig:results}
\end{figure*}

\begingroup

\begin{table*}
\caption{\label{tab:table}Values for $\xi$, $I$, $g_I^{\text{NP}}$ \& specific differences ($D_{21}$ for helium \& hydrogen, $\triangle' E$ for the rest) for different elements. TW indicates uncertainty estimates derived in this work (see the Appendix).}

\setcellgapes{1.5pt}\makegapedcells

\begin{center}
\begin{ruledtabular}
\begin{tabular*}{0.9\linewidth}{@{\extracolsep{\fill}}lcccdccl}
Element & $I$& \multicolumn{2}{c}{$|g_I^{\text{NP}}|$} &\multicolumn{1}{c}{$\xi$} & \multicolumn{1}{c}{Theory (meV)} & \multicolumn{1}{c}{Experiment (meV)} &Refs.\\
\\
\Xhline{.6pt}
 & & Pseudosc. & Vector & & & & \\
\Xhline{.6pt}
${}^{1}$H   &1/2 & 2& 2 & 0.125& $2.02458(10)\times 10^{-7}$ & $2.02479(28)\times 10^{-7}$ &\cite{Cong:2024nat} \\
${}^{3}$He   & 1/2 & 2& 2&0.125& $-
4.9218(6) \times 10^{-6}$  & $-4.92136(29) \times 10^{-6}$  & \cite{KARSHENBOIM20051}\\
${}^{9}$Be   & 3/2& $2/3$& $4/3$ &0.04881891046 & $-1.224(15)\times 10^{-6}$& $-1.13581526(5)\times 10^{-6}$  &\cite{Dickopf_2024} \\
${}^{131}$Xe & 3/2 & $2/5$& 4/5& 0.13660 & $\triangle' E_{\text{Xe}} \pm 7 \times 10^{-5}$ &$\triangle' E_{\text{Xe}} \pm 1 \times 10^{-5}$ &\cite{Indelicato}, TW \\
${}^{133}$Cs & 7/2& $2/9$& $8/9$ &0.13745& $\triangle' E_{\text{Cs}} \pm 7 \times 10^{-5}$& $\triangle' E_{\text{Cs}} \pm 1 \times 10^{-5}$ & \cite{Indelicato}, TW\\
${}^{209}$Bi & 9/2 & $2/9$ & $10/9$ & 0.16885 & $-61.043(35)$ & $-61.012(26)$ & \cite{10.1038/ncomms15484, PhysRevLett.86.3959} \\
\end{tabular*}
\end{ruledtabular}
\end{center}
\end{table*}
\endgroup

{\it Results.\ }-- We follow an approach similar to Ref.~\cite{Cong:2024nat}, determining the 95\% confidence interval (C.I.) for the difference between theory and experiment.
The Gaussian parameters are
\begin{eqnarray}
    \bar{x} &=& |\text{Theory} - \text{Experiment}| \nonumber \\
    \sigma &=& \sqrt{\sigma_{\text{theo}}^2 + \sigma_{\text{exp}}^2}.
\end{eqnarray}
To set our constraints we use 
the $95 \%$ C.I.\ bound 
$\delta E = \delta' E + \bar{x}$,
with
$\delta'E$ defined by
\begin{equation}
    95\% = \frac{1}{\sqrt{2\pi \sigma}} \int_{-\delta' E+\bar{x}}^{\delta' E+\bar{x}} e^{\frac{-(x-\bar{x})^2}{2\sigma}}\,\text{d}x,
\end{equation}
yielding conservative bounds.
For caesium and xenon projections we set $\bar{x}=0$ and use estimated uncertainties. Bounds then follow from
\begin{equation} \label{eq:constr}
    |\triangle E_{\text{Li-like}} - \xi \triangle E_{\text{H-like}}| \leq \delta E.
\end{equation}
We also include possible new-boson contributions to the nuclear $g$-factor in the corresponding measurement via the Zeeman splitting of hyperfine levels~\cite{PhysRevLett.107.043004,PhysRevA.85.022512}. Since the HFS and the nuclear $g$-factor measurement form a non-degenerate system, the two can be disentangled. This is outlined in the Appendix. We find that taking this into account reproduces the constraints derived from Eq.\ \eqref{eq:constr}, shown in Fig. \ref{fig:results}. 

\par

Future Cs measurements promise to extend the sensitivity to the electron–proton couplings, by up to a factor of 2-2.5 for pseudoscalars, and an order of magnitude for vectors for higher masses, see the Appendix for the
discussion regarding required uncertainties of theory and experiment. Using Be already matches or improves the sensitivity to pseudoscalar 
couplings with neutrons by up to a factor of $2$, depending on the nuclear model. These constraints supersede astrophysical exclusions in Ref.\ \cite{Cong:2024qly}, which stop at $m_{\phi} \gtrsim 10^4 \, \text{eV}$. We note the expected feature that heavier elements provide better bounds at higher masses due to the lower mean distance between the electrons and the nucleus. However, this is
tempered by reduced theoretical and experimental precision. At low masses, differences between bounds cannot be attributed to sensitivities alone. We also find that the strength of the constraint depends on the scaling of the specific difference, i.e., if the specific difference due to new physics scales similarly to the BW effect, the resulting bounds are less stringent. This occurs for hydrogen and helium, where the scaling is almost identical. This is the suppression we mentioned in the introduction, indicating that heavier elements can indeed be better suited to derive such constraints despite the reduced sensitivities due to nuclear effects.\par
Said suppression can be attributed to the systems used to form the specific differences in helium and hydrogen not including interelectronic interactions, which enables us to consider the following ratio 
\begin{eqnarray}
    \frac{\int \text{d}r \, F_{2s}(r)G_{2s}(r) \beta(r) }{\int \text{d}r \, F_{1s}(r)G_{1s}(r) \beta(r)} &\overset{m_{\phi} \rightarrow 0}{\rightarrow}& \frac{1}{8} + \mathcal{O}\left((Z\alpha)^2\right),
\end{eqnarray}
implying (note that $\xi = 1/8$ for helium and hydrogen) that the new physics contributions in the hydrogen and helium specific differences cancel for low masses up to relativistic corrections. Note that this may also happen for heavier elements depending on the value of $\xi$. However, this is not the case for the systems we consider, for which we found that the missing suppression is not caused by computational inaccuracies. 

We note that astrophysical constraints on the electron coupling can be extended to higher masses using results from Ref.~\cite{Carenza_2021}.
However, the corresponding supernova bound on nucleons likely falls into the trapping regime, weakening the constraint (cf.\ Ref.~\cite{PhysRevD.109.023001}).

{\it Conclusion.\ }-- We derive bounds on new light bosons from hyperfine splittings in highly charged ions. We employ specific differences, which largely overcome theoretical limitations from the Bohr–Weisskopf effect. This demonstrates that future precision spectroscopy of mid to high-$Z$ ions can yield competitive bounds on the pseudoscalar and vector couplings, improving upon constraints from hydrogen and helium by up to a factor of $2$, and an order of magnitude, respectively.

This analysis of new physics induced atomic interactions can be extended beyond our work by considering additional interaction types and higher-order QED contributions.

{\it Acknowledgements.\ }-- C.\ Q.\ wishes to thank V.~A. Yerokhin, N.~S. Oreshkina, and L. P\"utter for insightful discussions. 
This work is supported by the Collaborative Research Centre 1225 funded by Deutsche Forschungsgemeinschaft (DFG, German Research Foundation)–Project No. 273811115–SFB 1225. This article comprises parts of the PhD thesis work of
C.\ Q.\ to be submitted to Heidelberg University.

\appendix
\onecolumngrid 
\section{Appendix}
\twocolumngrid
{ \it Considered systems, Hamiltonians \& electron wavefunctions.\ }-- We consider two different systems to obtain our bounds. The first system is the hydrogen-like ion, for which we use the following Hamiltonian
\begin{align}
    H_{\text{H-like}} = \pmb{\alpha}\cdot\pmb{p} + \beta m_{e} + V_{\rm C}(r),
\end{align}
where $\alpha^1, \alpha^2, \alpha^3$ and $\beta$ denote the corresponding Dirac matrices in the Dirac basis and $V_{\rm C}$ accounts for the Coulomb interaction. We model the Coulomb interaction as originating from a Fermi charge distribution \cite{Johnson, JOHNSON1985405}, i.e.,
\begin{equation}
    \rho_{\rm C}(r) = \frac{\rho_0}{1-e^{\frac{r-c}{a}}},
\end{equation}
where $c$ is the half-density radius, and $a$ can be given via the \textit{skin-thickness} $t$ as 
$a = 4 \ln 3 \cdot t$. This yields a constant value for $a$ regardless of the element under consideration.
The Hamiltonian for the lithium-like system is defined analogously as 
\begin{align}
    H_{\text{Li-like}} = \pmb{\alpha}\cdot\pmb{p} + \beta m_{e} + V(r).
\end{align}

Here, $V(r) = V_{\rm KS}(r)+V_{\rm C}(r)$ includes screening effects via the Kohn-Sham potential
\begin{eqnarray}
 V_{\rm KS}(r) &=& \alpha \int \frac{\rho(r')}{r_>} \, \text{d}r' - \frac{2\alpha}{3r} \left(r\rho(r)\frac{31}{82\pi^2}\right)^{1/3},\nonumber \\
\end{eqnarray}
with the probability density $\rho$ encoding the screening effects due to the electrons present in the ion. This density is given by 
\begin{equation}
    \rho (r) = 2(\tilde{F}_{1s}(r)^2 + \tilde{G}_{1s}(r)^2) + (\tilde{F}_{2s}(r)^2 + \tilde{G}_{2s}(r)^2).
\end{equation}
Here, the wavefunctions $\tilde{F},\tilde{G}$ were obtained self-consistently via the DHF procedure outlined in 
Ref.~\cite{Johnson}. Aside from the aforementioned DHF procedure, the corresponding wavefunctions are obtained by solving the respective Dirac equations, whose general form is as follows
\begin{align}
    \psi_{n,\kappa,m}(r) = 
    r^{-1}\begin{pmatrix}
    G_{n,\kappa}(r)\Omega_{\kappa,m}(\theta,\varphi)  \\
    iF_{n,\kappa}(r)\Omega_{-\kappa,m}(\theta,\varphi)
    \end{pmatrix}. \label{eq:wf}
\end{align}
Here, $F$ and $G$ denote the radial wavefunctions for different parity states, whereas $\Omega$ represents spherical spinors. The quantum number $n$ denotes the usual primary quantum number, while $\kappa = \pm(j+1/2) $ is the relativistic angular momentum number, and $m$ indicates the angular momentum projection. More information on these functions and their properties can be found in Ref.\ \cite{2006sham.book.....D}.\par

{\it The new-physics induced dipole moment \& comparisons with full-scale nuclear simulations.} -- As stated in the main text, the interactions facilitated by the inclusion of new pseudoscalars or vector bosons can be formulated as an interaction with a dipole
\begin{eqnarray}
    \pmb{\mu} = g^{l}\pmb{L} + g^{s}\pmb{S},
\end{eqnarray}
where we take $g^{s} = 2$ within our effective coupling framework, and indicate the coupling to the angular momentum as 
\begin{eqnarray}
    g^{l} =\begin{cases}
        1, \,\, \text{Vector} \\
        0, \,\, \text{Pseudoscalar}
    \end{cases}.
\end{eqnarray}
The choice of nuclear model now refers to a choice of states with which we calculate the induced nuclear $g_I^{\text{NP}}$-factor. In the Schmidt model, this yields
\begin{eqnarray}
    g_I^{\text{NP}} = I^{-1} \begin{cases}
        g^{l} + \frac{1}{2}g^s, \, &I = l+\frac{1}{2} \\[.5em]
        \frac{I}{I+1} (g^l(l+1)-\frac{1}{2}g^s), \, &I = l-\frac{1}{2}
    \end{cases},\quad
\end{eqnarray}
where $l$ is the orbital angular momentum of the valence nucleon. We can also compare with the literature \cite{Kimball_2015} by extracting their values for $\langle \pmb{L} \rangle$ and $\langle \pmb{S}\rangle$. This way, we correct for our overestimating of the coupling strength to the valence nucleon to obtain conservative constraints using up-to-date values. 
\par
{\it The effect of new physics on the measurement of the nuclear $g_I$-factor.} -- Determination of the Standard Model value of the specific differences requires the measured nuclear $g_I$-factor as an input. We therefore have to consider the possibility that this measurement, too, is affected by new physics contributions.
One possibility to measure the nuclear $g_I$-factor is via a measurement of the $g_F$-factor, of a combined system
    such as a hydrogen-like ion. The nuclear $g_I$-factor can then be extracted via a combined measurement of the Zeeman splitting of two hyperfine states (cf.\ Refs.\ ~\cite{PhysRevA.85.022512,PhysRevLett.107.043004})
    \begin{equation}
        \bar{g} = g_{F=I+1/2} + g_{F=I-1/2} = -2 \frac{m_e}{m_p} g_I (1-\sigma).
    \end{equation}
Here, the dimensionless term $\sigma$ contains spin-dependent shielding corrections to the bound electron $g$-factor. In the presence of a new boson, there are additional contributions to this quantity, given by the following diagram
    \begin{equation}
        \begin{fmffile}{shielding}
        \begin{fmfgraph*}(90,25)
        \fmfstraight
        \fmfleft{i1,i2} 
        \fmfright{o1,o2}
        \fmffreeze
        \fmf{phantom}{i1,v11,v12,o1}
        \fmf{dbl_plain}{i2,v21,v22,o2} 
        
        \fmffreeze
        \fmf{photon}{v21,v11}
        \fmfv{decor.shape=triangle,decor.size=.2cm}{v11}
        \fmf{dashes}{v22,v12}
        \fmfv{decor.shape=square,decor.size=.15cm}{v12}
    \end{fmfgraph*}
    \end{fmffile},
    \end{equation}  
    where the wavy line terminated by a triangle denotes an insertion of the Zeemann potential induced by an external magnetic field \footnote{The distance between the magnetic field source and the ion is large compared to the range of the NP potential, $1/m_{\phi/Z}$, relevant for our constraints. Thus, we can neglect the effect that new bosons may have on this potential.}
    \begin{equation}
        V_{\rm Zee} (r) = \frac{|e|}{2} \pmb{B}\cdot (\pmb{r} \times \pmb{\alpha}).
    \end{equation}
    In analogy to the QED case presented in Refs.\ ~\cite{PhysRevLett.107.043004,PhysRevA.85.022512}, this diagram induces an additional energy shift, given by 
    \begin{eqnarray}
        \triangle E &=& 2 \sum_{n \neq a}\sum_{\substack{M_I,M_I'\\M_I'',M_I'''}}\sum_{\substack{m_a,m_a'\\m_n,m_n'}} C^{FM_F}_{j_am_a,I M_I}C^{FM_F}_{j_am_a',IM_I'}\nonumber \\&\times& C^{FM_F}_{j_nm_n,I M_I''}C^{FM_F}_{j_nm_n',IM_I'''}\frac{1}{\varepsilon_a -\epsilon_n}  \nonumber \\[1em] &\times& \bra{I M_I'}\bra{a} V_{\rm Zee}\ket{n}\ket{I M_I''}\nonumber \\[1em]
        &\times& \bra{I M_I'''}\bra{n} V_{\phi/Z}\ket{a} \ket{I M_I}.
    \end{eqnarray}
    For the measurements in question, we consider the case $j_a = 1/2$. The nuclear prefactors can then be factored out, yielding the following expression for the shielding contribution
    \begin{equation}
        \sigma_{p} = \frac{\triangle E}{M_F B \frac{\langle \pmb{I}\cdot \pmb{F}\rangle}{ F(F+1)}},
    \end{equation}
    with the usual angular factor known from hyperfine splitting calculations 
    \begin{equation}
        \langle \pmb{I}\cdot \pmb{F}\rangle = (F(F+1)-I(I+1)-j(j+1))/2,
    \end{equation}
    where $j$ denotes the electron spin, $I$ the nuclear spin, and $F$ stands for the spin of the combined system.
    From this, we can see that the equation used to determine the nuclear $g$-factor is modified as 
    \begin{equation}
        \bar{g}_p = \bar{g} + 2 \frac{m_e}{m_p} \sigma_p.
    \end{equation}
    Similarly, we can write down the SM and new physics contributions to the specific difference of hyperfine splitting intervals. Considering both at the same time, we can connect the variation of both observables with the following system of equations
    \begin{eqnarray}
        \begin{pmatrix}
            \delta (\triangle' E)\\
            \delta \bar{g}
        \end{pmatrix}
        =
        \begin{pmatrix}
            \frac{\partial \triangle' E}{\partial g_I} & \frac{\partial \triangle' E}{\partial g_eg_N} \\[1em]
            \frac{\partial \bar{g}}{\partial g_I} & \frac{\partial \bar{g}}{\partial g_eg_N}
        \end{pmatrix}
        \begin{pmatrix}
            \delta g_I \\
            \delta( g_e g_N)
        \end{pmatrix},
    \end{eqnarray}
    where $g_N \in \{g_n,g_p\}$ denotes either the neutron or proton coupling of the pseudoscalar (the treatment is analogous for vector bosons). We find that the matrix in the above equation is invertible within the mass range in which we derive our bounds, meaning we can disentangle the two measurements. Our bounds are then also obtained by solving the above system of equations (see also~\cite{Jaeckel_2010}), i.e., 
    \begin{equation}
        |\delta (g_eg_p) |\leq \frac{\left|\frac{\partial \triangle' E}{\partial g_I} \delta\bar{g}\right| + \left|\frac{\partial \bar{g}}{\partial g_I}\delta (\triangle' E)\right|}{\left|\frac{\partial \triangle' E}{\partial g_I}\frac{\partial \bar{g}}{\partial g_eg_N}-\frac{\partial \triangle' E}{\partial g_eg_N}\frac{\partial \bar{g}}{\partial g_I}\right|}\,.
    \end{equation}
    We find that deriving the bounds this way reproduces the ``naive`` bounds obtained from Eq.~\eqref{eq:constr}, implying that we can indeed resolve the two measurements via the solution of the coupled equations.
    Thus, we conclude that the bounds derived from Eq.\ \eqref{eq:constr} apply, even when taking the possible variation of the nuclear $g$-factor into account. 
The treatment for other types of measurements, such as NMR, is less straightforward, and it is not as easily established that we can disentangle them from the HFS measurements. Since this is the case for bismuth, our constraints are only a projection assuming the above treatment gives the same bounds, i.e., a Zeeman measurement yields the same result as the NMR measurement.
{\it Connection to the non-relativistic result.\ }-- Although our treatment of the dynamics of the bound electrons is fully relativistic, it is useful to see if our calculations reduce to the known non-relativistic results found in the literature \cite{Carenza_2021, Moody:1984ba}.
To this end, we investigate the non-relativistic case as follows: for brevity's sake, consider simplified wave functions of the type
\begin{equation}
    \psi = \begin{pmatrix}
        \varphi_1 \\
        \varphi_2
    \end{pmatrix},
\end{equation}
where $\varphi_1$ denotes the large, and $\varphi_2$ the small component of the \emph{full} Dirac spinor. 
In the non-relativistic limit, i.e., $E \approx m_e$ and $Z\alpha \ll 1$, one of the coupled Dirac-Coulomb equations simplifies to 
\begin{equation}
    -i\pmb{\sigma}_e \nabla \varphi_1 -2m_e\varphi_2 \approx 0.
\end{equation}
This enables us to rewrite the full HFS matrix element using partial integration 
\begin{eqnarray}
    \int \text{d}^3\pmb{r} \,[ -\varphi_2^{\dagger} \tilde{V}\varphi_1 + h.c.] = \frac{-i}{2m_e} \int \text{d}^3\pmb{r} \, \varphi_1^{\dagger}(\pmb{\sigma}_e\nabla\tilde{V})\varphi_1, \nonumber \\
\end{eqnarray}
where we used that the integrand vanishes on the boundary at infinity, and defined the following modified potential term 
\begin{eqnarray}
    \tilde{V} 
    &=& -i \frac{g_Ng_e}{8\pi m_N}(\pmb{\sigma}_N\cdot \nabla) \left( \frac{e^{-m_{\phi}r}}{r} \right).
\end{eqnarray}
Using this definition, we see that the procedure outlined above yields a new potential term within the integral. This new potential is given by \cite{Fadeev_2019} 
\begin{eqnarray}
    i(\pmb{\sigma}_e\cdot\nabla)\tilde{V} &=& \frac{g_Ng_e}{16 \pi m_e m_N} (\pmb{\sigma}_e\cdot \nabla)(\pmb{\sigma}_N\cdot \nabla)\left( \frac{e^{-m_{\phi}r}}{r} \right). \nonumber \\
\end{eqnarray}
This matches the non-relativistic pseudoscalar-mediated potential usually found in the literature (cf.\ Ref.\ ~\cite{Fadeev_2019, Moody:1984ba}), confirming that our relativistic formulation is consistent with the non-relativistic approach. This treatment can be carried out analogously for vector bosons.

{\it Photon exchange \& counter term diagrams. }-- Here, we give the algebraic expressions for the first-order interelectronic interaction diagrams. For a more thorough treatment and discussion of these contributions, one may refer to Refs.\ ~\cite{PhysRevResearch.2.013364, Shabaev_2002}.\par
The energy shift due to the one-photon exchange diagram is given by the following
expression

\begin{widetext}
\begin{eqnarray}
    \triangle E^{(1)}_{(1s)^2 2s,\gamma} \!\!&=&\!\! \sum_b \sum_{m_a,M_I} \sum_{m_a'M_I'} C^{FM_F}_{j_am_a,IM_I}C^{FM_F}_{j_am_a',IM_I'}\bra{IM_I'}  \\
    \!\!&\times&\!\! \bigg[\sum_P (-1)^P (\bra{\xi_{b|PaPb}}V_\phi\ket{a} + \bra{\xi_{a|PbPa}}V_{\phi} \ket{b}) - \frac{1}{2}(\bra{a}V_{\phi}\ket{a} + \bra{b}V_{\phi}\ket{b})\bra{ab}I'(\varepsilon_a-\varepsilon_b)\ket{ba}\bigg] \ket{IM_I},\nonumber 
\end{eqnarray}
with the following identities 
\begin{eqnarray}
    \ket{\xi_{a|PbPa}} = \sum_{n} \ket{n} \frac{\bra{na}I(\triangle)\ket{PbPa}}{\varepsilon_b-\varepsilon_n} \quad \text{and} \quad \ket{\xi_{b|PaPb}} = \sum_{n} \ket{n} \frac{\bra{nb}I(\triangle)\ket{PaPb}}{\varepsilon_a-\varepsilon_n},
\end{eqnarray}
\end{widetext}
where $P$ denotes the permutation operator, $\varepsilon$ stands for the eigenenergies of the indexed states, and $\ket{a} = \ket{j_am_a}$ represent electron states, with $a$ being the valence ($2s$) state, and $b$ denoting the core ($1s$) states, respectively. The states $\ket{IM_I}$ refer to the nuclear angular momentum states, which we couple to the valence electrons via the addition of angular momenta. Moreover, we denote the photon exchange operator as $I(\omega)$. It is defined as
\begin{equation}
    I(\omega,r_{12}) = \alpha \frac{1-\pmb{\alpha}_1\pmb{\alpha}_2}{r_{12}} e^{i\sqrt{\omega+i0}\, r_{12}},
\end{equation}
in the Feynman gauge,
with the arguments being $\triangle = \varepsilon_a - \varepsilon_{Pa}$, and $r_{12} = |\pmb{r}_1 - \pmb{r}_2|$.
Finally, the contribution due to the counter term is given by 
\begin{equation}
    E^{(1)}_{(1s)^2 2s,\text{counter}} = 2\sum_{n} \bra{a}V_{\phi}\ket{n} \frac{\bra{n} V_C(r) - V(r) \ket{a}}{\varepsilon_a - \varepsilon_n},
\end{equation}
where the factor of two arises as a symmetry factor due to the possible permutations of the diagram. We further suppressed the coupling to angular momentum in this contribution. This needs to be carried out analogously to the previous diagrams.

 \begin{figure*}[hbt!]
    \centering
    \includegraphics{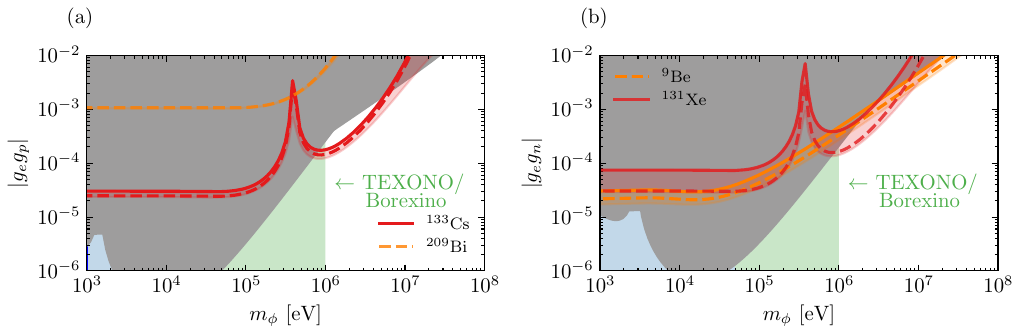}
    \includegraphics{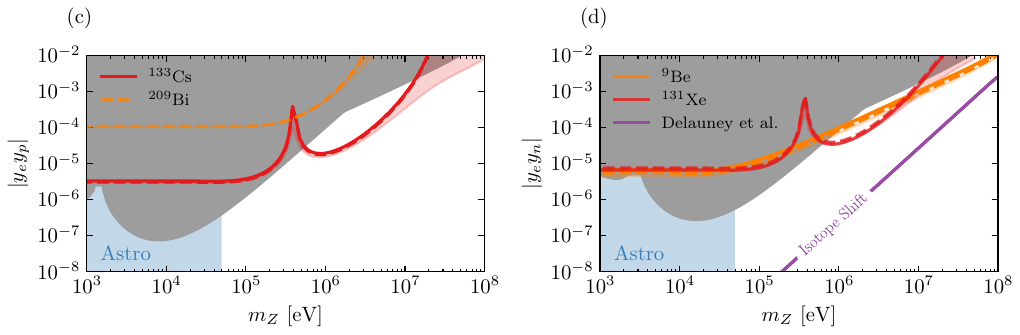}
    \caption{Version of our results that includes the error bands (shaded regions) on our constraints/sensitivity projections (solid lines) from our comparison with Ref.\ ~\cite{PhysRevResearch.2.013364}, as well as the errors arising from our choice of nuclear model for the coupling to the nucleus \cite{Kimball_2015}. Dashed lines indicate bounds derived within the one particle Schmidt model.}
    \label{fig:exclusions-error}
\end{figure*}

{\it Current \& projected sensitivities. }-- Aside from existing sensitivities derived from theoretical and experimental errors, we also give projected bounds. To this end, we spot possible improvements in theoretical calculations with regard to the realistically attainable relative precision, and apply them to the elements in question.
Up to now, a complete calculation of the specific difference for heavy HCI is only available for $^{209}\text{Bi}$
~\cite{10.1038/ncomms15484}. The relative uncertainties on the theory values are rather similar for HCI in the range $50 \leq Z \leq 83$~\cite{PhysRevA.56.252,PhysRevA.57.149}. In this range, the main differences are the effect of the nuclear charge distributions (Breit-Rosenthal (BR) effect ~\cite{PhysRev.41.459,PhysRev.76.1310}) and the nuclear magnetization distributions (BW effect ~\cite{PhysRev.77.94}). Their influence increases up to a factor of 10 for heavy HCI such as $^{209}\text{Bi}^{82+,80+}$ compared to, e.g., $^{133}\text{Cs}^{54+,52+}$. \par 
We use the published calculations for the specific difference of $^{209}\text{Bi}$ and the quoted uncertainties shown in Ref.~\cite{10.1038/ncomms15484} to estimate the possible theory uncertainty for the specific difference of $^{133}\text{Cs}$, $^{115}\text{In}$, and $^{131}\text{Xe}$, which are rather close in nuclear charge. We, therefore, assume the same uncertainties for those isotopes, starting from the one we derive for caesium.
In Ref.\ ~\cite{10.1038/ncomms15484}, five sources of uncertainties are listed for the theory value. All have a similar size. 

First, there is the uncertainty due to the higher order interelectronic interactions $~1/Z^3$ with a relative precision of $5 \times 10^{-5}$. There have been new calculations, achieving relative precision below $10^{-6}$ for $^{133}\text{Cs}^{52+}$~\cite{PhysRevResearch.2.013364}. Therefore, the remaining uncertainty due to the interlectronic interaction is not a relevant contribution to the overall uncertainty, at least relatively speaking.  \par
 Contributions from screened QED calculations have a relative uncertainty of $3 \times 10^{-5}$~\cite{PhysRevA.85.022510}. The dominant uncertainty is given by one  diagram regarding photon exchange with an additional bound-electron loop that mediates the hyperfine interaction, that has not been evaluated yet. We expect it to be possible to calculate this remaining diagram, and thereby improve the relative precision of the screened QED contributions by a factor of 3 for the HFS in $^{133}\text{Cs}^{52+}$. \par
Although the specific difference is constructed in a way that ensures suppression of the BW effect, there is still a remaining contribution due to the BW effect influencing interelectronic interaction contributions~\cite{doi:10.1139/p00-060}. This part is not completely canceled in the specific difference, resulting in a corresponding uncertainty. The contribution of the BW effect in $^{133}\text{Cs}^{54+,52+}$ is a factor of 7 smaller compared to the one in  $^{209}\text{Bi}^{82+,80+}$~\cite{PhysRevA.56.252,PhysRevA.57.149}. Additionally, the attributed uncertainty for the BW effect in $^{133}\text{Cs}^{54+,52+}$ is a factor of three smaller compared to the one for $^{209}\text{Bi}^{82+,80+}$. Therefore, we estimate the remaining uncertainty in the specific difference to be also a factor of three smaller. The estimated relative uncertainty regarding the remaining  BW effect in the specific difference in caesium then amounts to $2 \times 10^{-5}$.  

The remaining uncertainty due to nuclear polarization has been improved by more than a factor of 50 to a relative precision of $4 \times 10^{-7}$ for the specific difference of $^{209}\text{Bi}$ ~\cite{PhysRevLett.113.023002}. We assume a similar relative uncertainty for $ ^{133}\text{Cs}$ as well.

The nuclear magnetic moment of the nucleus may be determined via a sub ppb precision measurement of the $g_F$ factor in the ground and the excited state~\cite{PhysRevA.78.032517,Herfurth_2015,Vogel_2015}. Recently, the bound electron $g_j$ factor of  $^{118}\text{Sn}^{49+}$ has been successfully measured with a relative precision $5 \times 10^{-10}$~\cite{Morgner2023}. With further realistic improvements on the experimental side, we expect a possible precision of the measured $g_F$ factor of $1 \times 10^{-10}$. Together with the theoretical shielding calculations~\cite{PhysRevA.70.032105,PhysRevLett.107.043004} this may lead to a nuclear magnetic moment for $^{133}\text{Cs}$ with a relative precision below $5 \times 10^{-6}$, which scales similar for the specific difference of $^{133}\text{Cs}$ as well. 

Combining the five estimated uncertainties for the theory value of $\Delta ' E \left(^{133}\text{Cs}\right)$, we find that improvements regarding the theoretical precision can be made up to a relative uncertainty of $2 \times 10^{-5}$ mainly given by the screened QED uncertainty together with the remaining BW uncertainty.

Finally, a possible measurement of $E_{\text{HFS}} \left(^{133}\text{Cs}^{54+} \right)$ and  $E_{\text{HFS}} \left( ^{133}\text{Cs}^{52+} \right)$ to MHz precision will lead to $\Delta ' E \left(^{133}\text{Cs}\right)$ with a relative precision of $3 \times 10^{-6}$ on the experimental side. This might eventually be feasible and achievable in the future, since both wavelengths ($1.9\,\mu$m and $14\,\mu$m) are accessible by adequate lasers.

We assume similar relative theory uncertainties for $\Delta ' E \left(^{115}\text{In}\right)$ and $\Delta ' E \left(^{131}\text{Xe}\right)$, and a similar experimental precision for $\Delta ' E \left(^{115}\text{In}\right)$, since both wavelengths ($1.4\,\mu$m and $11\,\mu$m) are accessible by adequate lasers, too. This is unfortunately not the case for $\Delta ' E \left(^{131}\text{Xe}\right)$, where the corresponding wavelength of the lithium-like HFS splitting is larger than $20\,\mu$m strongly limiting accessible lasers. Therefore, to achieve a $\sim 1\,{\rm{MHz}}$ precision for the two HFS energy measurements remains a challenging task. 

It is very important to experimentally determine the nuclear magnetic moment to a relative
precision of $5\times10^{-6}$ and the hyperfine energy splitting to a relative precision of around $3\times10^{-6}$ to extract the constrains on pseudoscalar and vector bosons based on the specific difference of $^{133}$Cs
as plotted in Fig.\ \ref{fig:exclusions-error}.
For the two HFS transitions in $^{133}$Cs this corresponds to an experimental precision of tens of MHZ for $\triangle E_{\text{HFS}}(^{133}\text{Cs}^{54+})$ and 1 MHz for the $\triangle E_{\text{HFS}}(^{133}\text{Cs}^{52+})$. This is technically feasible,
since both wavelengths (1.9 $\mu$m and 14 $\mu$m) are accessible by adequate lasers. On the one hand this precision is a factor of around three more precise compared to the current best measurements carried out on an ion storage ring \cite{PhysRevLett.120.093001, 10.1038/ncomms15484}, which are limited by the high-voltage measurement required for the precise energy determination. On the other
hand, it is possible to reach the tens of MHz precision for cryogenically cold trapped single highly charged ions \cite{PhysRevLett.123.123001}. Here the uncertainties due to the ion energy are in the range of single kHz.
Importantly, it is not sufficient to determine the nuclear magnetic moment based on nuclear magnetic resonance measurements of macroscopic samples. Here the interaction of the new bosons is very hard to quantify. The largest systematic uncertainty on the
nuclear magnetic moment of $^{133}$Cs via a direct $g_F$ factor measurement of a single hydrogen-like caesium ion will be below $1\times10^{-6}$ and is due to the interaction of the ion with the detection system.

{\it Error estimates on our constraints. }-- Although the methods outlined in this work should provide quite accurate results, some care is needed, as the specific differences also lead to subtractions in the signal calculation. Hence, we need to test the sensitivity of our signal calculation to approximation errors.

In principle, further theoretical improvements are possible by accounting for higher orders of interelectronic interactions in the lithium-like system. We can, however, obtain a rough estimate on the uncertainties caused by not including these contributions by comparing our results for the SM hyperfine splitting in lithium-like systems with the ones given in Ref.~\cite{PhysRevResearch.2.013364}. 
 We assume that our calculations' error is due to missing diagrams and inaccuracies in our wavefunctions and carry this relative uncertainty over into our calculations for the pseudoscalar and vector-induced hyperfine splitting. Of course, this also implicitly assumes that the relative error is independent of the mass scale in the potential.
 Since the deviations from literature vary only slowly with $Z$, we apply the same relative error for elements with a sufficiently similar nuclear charge. We find a relative uncertainty of about $0.1\%$ for mid to high-$Z$ elements, and $1\%$ for low-$Z$ elements.\par
We further include modelling uncertainties arising from the use of the single particle Schmidt-model for caesium, xenon, beryllium, and xenon by following the comparison of nuclear models outlined in Ref.\ \cite{Kimball_2015}.
 In the main text, we show the most conservative constraints after incorporating these uncertainties. The corresponding error bands are shown in Fig.\ ~\ref{fig:exclusions-error}.

\bibliography{ref}

\end{document}